# Interval Type-2 Enhanced Possibilistic Fuzzy C-Means Clustering for Gene Expression Data Analysis

**Shahabeddin Sotudian[a], Mohammad Hossein Fazel Zarandi[b,c]**

[a] Department of Electrical and Computer Engineering, Division of Systems Engineering, Boston University, Boston, USA
[b] Knowledge/Intelligent Systems Laboratory, Department of Mechanical and Industrial Engineering, University of Toronto, Toronto, Canada
[c] Department of Industrial Engineering, Amirkabir University of Technology, Tehran, Iran

sotudian@bu.edu, zarandi@aut.ac.ir

**Abstract**

Both FCM and PCM clustering methods have been widely applied to pattern recognition and data clustering. Nevertheless, FCM is sensitive to noise and PCM occasionally generates coincident clusters. PFCM is an extension of the PCM model by combining FCM and PCM, but this method still suffers from the weaknesses of PCM and FCM. In the current paper, the weaknesses of the PFCM algorithm are corrected and the enhanced possibilistic fuzzy c-means (EPFCM) clustering algorithm is presented. EPFCM can still be sensitive to noise. Therefore, we propose an interval type-2 enhanced possibilistic fuzzy c-means (IT2EPFCM) clustering method by utilizing two fuzzifiers $(m_1, m_2)$ for fuzzy memberships and two fuzzifiers $(\theta_1, \theta_2)$ for possibilistic typicalities. Our computational results show the superiority of the proposed approaches compared with several state-of-the-art techniques in the literature. Finally, the proposed methods are implemented for analyzing microarray gene expression data.

## 1. Introduction

Machine learning has shown promising results in a wide range of applications (Hao et al., 2020, p. 19; Sotudian & Paschalidis, 2021; Zandigohar et al., 2020) in recent years. As an unsupervised ML technique, clustering is a pattern classification method that forms clusters by grouping data samples into homogeneous clusters. Among various approaches for clustering, fuzzy c-means clustering (FCM) is one of the most popular in the real-world application due to its fast convergence, simplicity, and easy implementation(Dagher, 2018). That is why it has been extensively applied in various fields such as engineering and medical sciences (Sotudian et al., 2016, 2021; Su et al., 2021). Unfortunately, FCM-type algorithms share the problem of sensitivity to noise and outliers due to low values of membership associated with noisy points. In this regard, many variations of this algorithm have been developed to address this issue (Cheng et al., 2019; Ding & Fu, 2016). Among all the variants of FCM, one particularly important research is that of Krishnapuram and Keller (R. Krishnapuram & J. M. Keller, 1993). They proposed and later improved the possibilistic c-means algorithm (PCM). They relaxed the probabilistic constraint imposed on the membership value in the FCM algorithm to avoid the effect of outliers on the clustering performance. Even though PCM solves FCM's sensitivity to noise and outliers, it has the disadvantage of generating coincident clusters (Zhou & Hung, 2007).

To overcome the shortcomings of PCM, Pal et al. proposed two hybridizations of FCM and PCM. In 1997, they proposed fuzzy possibilistic c-means (FPCM) clustering for simultaneously constructing memberships and typicalities. However, FPCM produces unrealistic typicality values for large data sets. In 2005, they proposed the possibilistic fuzzy c-means (PFCM) clustering method that they believe can avoid various problems of PCM, FCM, and FPCM (Pal et al., 2005). Although PFCM enhances the clustering results and alleviates the coincident-clusters problem, it is not able to completely solve the issue as we will demonstrate in the next section.

Furthermore, the clustering algorithms mentioned above cannot properly handle the uncertainty in real-world problems. As a consequence, these algorithms have been improved using type-2 fuzzy logic techniques. Type-2 fuzzy sets (T2 FSs) can model different forms of uncertainty by their fuzzy membership values rather than crisp ones (Hwang & Rhee, 2007). Although the computational complexity is increased by the employment of T2 FSs, it is a small price to pay for obtaining satisfactory results. Moreover, using T2 FSs, we can enhance the algorithms' resilience against noisy points and outliers. In type-2 fuzzy clustering algorithms, the degree of belonging to the cluster representatives is represented as either an interval-valued type-2 fuzzy set (IT2 FS) or as a general type-2 fuzzy set (GT2 FS). In the literature, type-2 fuzzy clustering models have mainly defined using IT2 FSs because its computational complexity is less than the computational complexity of GT2 FSs (Melin & Castillo, 2014).

In the current paper, we will focus on PFCM. More specifically, we will modify the conventional PFCM algorithm so that the proposed clustering method be able to overcome various problems of PCM, FCM, FPCM, and PFCM. With this in mind, the main contributions of our paper are as follows:

- we will enhance the conventional PFCM algorithm, giving rise to a modified algorithm which will be called Enhanced Possibilistic Fuzzy C-Means (EPFCM) clustering. This method can not only overcome the negative impact of the noisy points efficiently but also reduce closely located or coincident clusters. Moreover, EPFCM does not produce unrealistic typicality values for large data sets which we usually see in FPCM-based clustering methods. These preferable characteristics of EPFCM make it satisfactory to be the basic model in our study.

- Higher-order fuzzy clustering algorithms have been demonstrated to be very capable to deal with the high levels of uncertainties that exist in most of the real-world applications. We will incorporate interval type-2 fuzzy sets (IT2 FSs) into the EPFCM algorithm to enhance the flexibility of the model for handling uncertainty and vagueness. The uncertainty associated with fuzzifiers $m$ and $\theta$ is considered in the proposed algorithm. In this way, we also enhance the ability of our model to overcome the coincident-clusters problem.

- Finally, the proposed algorithms are applied in the analysis of microarray gene expression data. Our results show that the proposed methods are more robust to outliers and initializations and can produce more accurate clustering results.

Based on the above discussions, the remainder of this paper is organized as follows. In Section 2, the problem statement and motivations will be presented. In Section 3, the enhanced possibilistic fuzzy c-means algorithm is formulated. In Section 4, the proposed interval type-2 enhanced possibilistic fuzzy c-means algorithm and its properties are presented. Several examples and comparisons showing the validity of our proposed methods will be presented in Section 5. Section 6 is devoted to the application of the proposed methods in microarray gene expression data clustering. Finally, conclusions and future work are presented in Section 7.

## 2. Problem statement and motivations

As we discussed briefly in the introduction section, to avoid the tendency to produce coincident clusters, possibilistic fuzzy c-means (PFCM) was proposed by Pal et al. in 2005. PFCM uses the membership and typicality aspects of FCM and PCM and minimizes the following optimization problem (Pal et al., 2005):

$$\min \left\{ J_{PFCM}(U, T, V; X) = \sum_{i=1}^{c} \sum_{j=1}^{n} (C_{FCM}(u_{ij})^m + C_{PCM}(t_{ij})^\theta) \|x_j - v_i\|_A^2 + \sum_{i=1}^{c} \eta_i \sum_{j=1}^{n} (1 - t_{ij})^\theta \right\}, \quad (1)$$

where $X$ is a set of all data points, $T$ is the typicality matrix, $t_{ij}$ is taken as the typicality of $x_j$ in the $i^{th}$ cluster of $X$. $V$ is a vector of cluster centers, $U$ is the partition matrix, and $u_{ij}$ is taken as the membership of $x_j$ in the $i^{th}$ cluster of $X$. $m > 1$, $\eta > 0$, and $\theta > 1$ are user-defined constants, $x$ represents a data point, $\|x\|_A = \sqrt{x^T A x}$ is any inner product norm, $n$ is the number of data points, and $c$ is the number of cluster centers. The constants $C_{FCM}$ and $C_{PCM}$ define the relative importance of fuzzy membership and typicality values in the objective function, respectively.

Pal et al. believed that PFCM can simultaneously exhibit the invulnerability of FCM to the coincident-cluster problem and the robustness of PCM to outliers. However, the coincident-clusters problem is still observed in the experimental result of this algorithm. To shed more light on the issue, consider the dataset $D_1$ in Figure 1 (a) which consists of 1550 data points and five clusters. We also added two clusters of outliers far from other points (each cluster contains 25 data points). We know when we use a higher value for $C_{FCM}$ compared to $C_{PCM}$, we force PFCM to behave more like FCM than PCM. In Figure 1 (b), we can see the clustering result for $C_{FCM} = 0.8$ and $C_{PCM} = 0.2$ (For this example, we set $\theta = 4$ and $m = 2$). Evidently, we can see that the coincident-clusters problem happened again. Despite the fact that PFCM enhances the clustering results and alleviates the coincident-clusters problem, it is not able to completely solve the issue as we will demonstrate in the example.

Hence, it can be concluded that the existing PFCM method cannot provide satisfactory performance for real-world applications. Thus, we need a better way to cope with this problem. In the next section, we will present EPFCM clustering which is a modified version of PFCM. It will be demonstrated that EPFCM can overcome various problems of PCM, FCM, FPCM, and PFCM.

## 3. Enhanced Possibilistic Fuzzy C-Means

As stated in the previous section, the existing PFCM algorithm still suffers from the coincident-clusters problem. This is due to the structural weakness that exists in the PCM part of this model. Inspired by the PCM algorithm of Krishnapuram and Keller(Krishnapuram & Keller, 1996), we modify the PCM part of PFCM algorithm. The objective function of the enhanced possibilistic fuzzy c-means (EPFCM) clustering is presented as follows:

$$\min \left\{ J_{EPFCM}(U,T,V;X) = \sum_{i=1}^{c} \sum_{j=1}^{n} (C_{FCM}(u_{ij})^m + C_{PCM}(t_{ij})^\theta) \|x_j - v_i\|_A^2 + \sum_{i=1}^{c} \eta_i \sum_{j=1}^{n} \sqrt[\theta]{t_{ij}} \left( \log\left(\sqrt[\theta]{t_{ij}}\right) - 1 \right) \right\}, \quad (2)$$

and the constraints are

$$0 \leq u_{ij} \leq 1, \quad \sum_{i=1}^{c} u_{ij} = 1, \quad \sum_{j=1}^{n} u_{ij} > 0, \tag{3}$$

$$0 \leq t_{ij} \leq 1, \quad \sum_{j=1}^{n} t_{ij} > 0.$$

Given that the parameters of this model are the same as PFCM, their redefining is avoided. The updating formulas for fuzzy membership, possibilistic typicality, and cluster centers can be written as follows:

$$u_{ij} = \left( \sum_{t=1}^{c} \left( \frac{D_{ij}}{D_{tj}} \right)^{\frac{2}{m-1}} \right)^{-1}, \tag{4}$$

$$t_{ij} = \left( \exp(-W(\mu)) \right)^{\frac{\theta}{\theta^2 - 1}}, \quad \mu = \frac{\theta^2 (\theta^2 - 1) C_{PCM} D_{ij}^{\ 2}}{\eta_i}, \tag{5}$$

$$v_i = \frac{\sum_{j=1}^{n} (C_{FCM}(u_{ij})^m + C_{PCM}(t_{ij})^{\theta}) x_j}{\sum_{j=1}^{n} (C_{FCM}(u_{ij})^m + C_{PCM}(t_{ij})^{\theta})} \quad , 1 \leq i \leq c, \tag{6}$$

where $D_{ij}$ represents the distance between $i^{th}$ cluster and $j^{th}$ data given by $D_{ij} = \|x_j - v_i\|$, and $W(.)$ is the Lambert W-function. Since $\mu$ is always non-negative, $W(\mu)$ has only one real value. Moreover, it should be noted that the value of $\eta_i$ can be calculated like PCM method using the following formula:

$$\eta_i = K \frac{\sum_{j=1}^{n} (u_{ij})^m \|x_j - v_i\|_A^2}{\sum_{j=1}^{n} (u_{ij})^m}, \tag{7}$$

where $K$ is a positive constant. The main steps of EPFCM clustering method are presented in Algorithm 1. Equation (2) and the constraint in Equations (3) can form a Lagrange equation as follows:

$$\mathcal{L} = \sum_{i=1}^{c} \sum_{j=1}^{n} (C_{FCM}(u_{ij})^m + C_{PCM}(t_{ij})^{\theta}) \|x_j - v_i\|_A^2 + \sum_{i=1}^{c} \eta_i \sum_{j=1}^{n} \left( \sqrt[\theta]{t_{ij}} (\log(\sqrt[\theta]{t_{ij}}) - 1) \right) - \sum_{j=1}^{n} \lambda_j \left( \sum_{i=1}^{c} (u_{ij} - 1) \right),$$

where $\lambda = [\lambda_1, ..., \lambda_n]$ is a Lagrange multiplier vector. Since the second part of $\mathcal{L}$ does not depend on $u_{ij}$ and $v_i$, the proofs of updating formulas for $u_{ij}$ and $v_i$ are exactly like original PFCM. Therefore, we do not repeat it (see (Krishnapuram & Keller, 1993; Pal et al., 2005) for details).

| **Algorithm 1:** EPFCM clustering method |
|---|
| **Initialization:** |
| Fix $c, K, m, \theta, \epsilon, C_{FCM}, C_{PCM}$ and $I_{Max}$ (maximum number of iterations); |
| Set iteration counter $t = 1$; |
| Initialize $U, T, \eta$ and $V$ randomly; |
| **While** ($\|U^{(t)} - U^{(t-1)}\| > \epsilon$ or $t < I_{Max}$) |
|     Update $U$ using (4). |
|     Update $T$ using (5). |
|     Update $V$ using (6). |
|     Update $\eta$ using (7). |
|     $t = t + 1$ |
| **End While** |

To find the first-order necessary conditions for optimality, the gradients of $\mathcal{L}$ with respect to $t_{ij}$ are set to zero:

$$\frac{\partial \mathcal{L}}{\partial t_{ij}} = \theta C_{PCM}(t_{ij})^{\theta-1} D_{ij}^{2} + \eta_i \left( \frac{(t_{ij})^{\frac{1-\theta}{\theta}} \log(\sqrt[\theta]{t_{ij}})}{\theta} \right) = 0,$$

which leads to

$$\frac{\theta^2 C_{PCM} D_{ij}^{2}}{\eta_i} (t_{ij})^{\theta - \frac{1}{\theta}} + \log(\sqrt[\theta]{t_{ij}}) = 0.$$

By solving this equation for $t_{ij}$, one can obtain

$$t_{ij} = \left( \exp\left( -W\left( \frac{\theta^2(\theta^2 - 1) C_{PCM} D_{ij}^{2}}{\eta_i} \right) \right) \right)^{\frac{\theta}{\theta^2 - 1}}$$

$(1 - t_{ij})^{\theta}$ in the original PFCM (Equation (1)) is monotonically decreasing function in [0,1]. Similarly, $\sqrt[\theta]{t_{ij}}(\log(\sqrt[\theta]{t_{ij}}) - 1)$ is also a monotonically decreasing function in [0,1]; therefore, it does not change the structure of PFCM algorithm. However, since the exponential function in the updating formula of EPFCM decays faster for large values of $D_{ij}^{2}$, EPFCM can manage to overcome the coincident-clusters problem of PFCM. As can be observed in Figure 1 (c), if we use the proposed algorithm for clustering $D_1$, EPFCM is able to solve the coincident-clusters problem of PFCM.

**Theorem.** The required condition of the convergence of the proposed algorithm is fulfilled when we have:

$$\lim_{t \to \infty} \|U^{(t)} - U^{(t-1)}\| = 0 \tag{8}$$

**Proof:** If we consider $J$ as the error function to be minimize, the iterative formula for $u_{ij}$ can be deduced from the Newton-Raphson method:

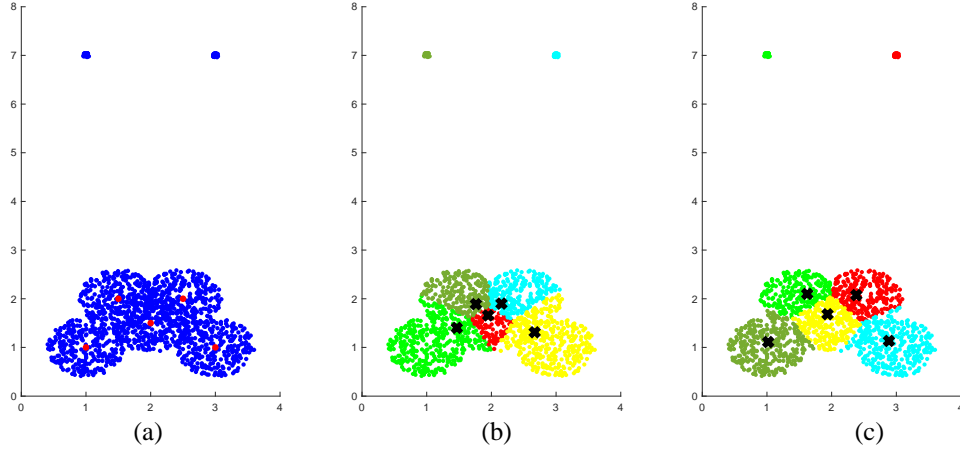

**Figure 1.** PFCM and coincident-cluster problem (a) Dataset $D_1$ and red dots are the center of clusters (b) The clustering result for PFCM (c) The clustering result for EPFCM.

$$u_{ij}^{(t)} = u_{ij}^{(t-1)} - \xi^{(t)} \left(J_{EPFCM}(u_{ij}, t_{ij}, v_i)^{(t-1)}\right) \left(\frac{\partial J_{EPFCM}(u_{ij}, t_{ij}, v_i)^{(t-1)}}{\partial u_{ij}}\right)^{-1},$$

where $\xi^{(t)}$ is a positive learning rate parameter, and $\frac{\partial J_{EPFCM}(u_{ij},t_{ij},v_i)^{(t-1)}}{\partial u_{ij}}$ is the gradient of $J_{EPFCM}$ with respect to $u_{ij}$ at $(t-1)$ iteration. By rewriting this equation for $U$, one can obtain:

$$U^{(t)} - U^{(t-1)} = -\xi^{(t)}\left(J_{EPFCM}(U,T,V)^{(t-1)}\right)\left(\frac{\partial J_{EPFCM}(U,T,V)^{(t-1)}}{\partial U}\right)^{-1}. \quad (9)$$

By putting Equation (9) in Equation (8) we have:

$$\lim_{t \to \infty} \|U^{(t)} - U^{(t-1)}\| = \lim_{t \to \infty} \left( \|\xi^{(t)}\| \left\|\partial J_{EPFCM}(u_{ij}, t_{ij}, v_i)^{(t-1)}\right\| \left\|\left(\frac{\partial J_{EPFCM}(u_{ij}, t_{ij}, v_i)^{(t-1)}}{\partial u_{ij}}\right)^{-1}\right\| \right).$$

Now, if we set $\xi^{(t)} = \dfrac{\bar{\xi}^{(t)}}{\left\|\partial J_{EPFCM}(u_{ij},t_{ij},v_i)^{(t-1)}\right\|\left\|\left(\frac{\partial J_{EPFCM}(u_{ij},t_{ij},v_i)^{(t-1)}}{\partial u_{ij}}\right)^{-1}\right\|}$, where $\bar{\xi}^{(t)} = \dfrac{\bar{\xi}_0^{(t)}}{t}$, $\bar{\xi}_0^{(t)}$ is a constant value and $\bar{\xi}^{(t)} \to 0$ when $t \to \infty$, the previous equation can be simplified as follows:

$$\lim_{t \to \infty}\|U^{(t)} - U^{(t-1)}\| = \lim_{t \to \infty}\left\|\bar{\xi}^{(t)}\right\| = 0.$$

Therefore, Equation (8) is proved and EPFCM converges.

## 4. Interval Type-2 Enhanced Possibilistic Fuzzy C-Means

The higher ability of IT2 FSs for modeling uncertainties is mostly attributed to their three-dimensional membership functions. T2 FSs have been used to handle the uncertainties in different domains where the performance of T1 FSs is not sufficiently good (Fazel Zarandi et al., 2018). Even though the computational complexity is increased by the employment of T2 FSs, but it is a small price to pay for obtaining satisfactory results (Hwang & Rhee, 2007). In this section, we improve the capabilities of the proposed model in uncertainty modeling using T2 FSs. Among the main approaches in the type-2 fuzzy clustering, the parameter uncertainty is a widely used method. This approach manages the amount of fuzziness of the final-partition by incorporating two values of fuzzifier $m$ to make a footprint of uncertainty (FOU) corresponding to the lower and upper interval memberships(Hwang & Rhee, 2007).

In the current study, we extend EPFCM to interval type-2 EPFCM using the Zexuan, et al. approach (Ji et al., 2014). For more detail on the concept of IT2 FSs and the techniques directly relevant to our work, please refer to(Ji et al., 2014). For representing the problem into IT2 FSs, we consider two fuzzifiers $(m_1, m_2)$ for fuzzy memberships and two fuzzifiers $(\theta_1, \theta_2)$ for possibilistic typicalities. These four fuzzifiers which represent different fuzzy degrees give four distinct objective functions to be minimized. Equation (10) shows these objective functions.

$$\begin{cases} J_{m_1,\theta_1}(U,T,V) = \sum_{i=1}^{c}\sum_{j=1}^{n}(C_{FCM}(u_{ij})^{m_1} + C_{PCM}(t_{ij})^{\theta_1})\|x_j - v_i\|_A^2 + \sum_{i=1}^{c}\eta_i\sum_{j=1}^{n}(t_{ij}^{\theta_1}\ln(t_{ij}^{\theta_1}) - t_{ij}^{\theta_1}) \\ J_{m_1,\theta_2}(U,T,V) = \sum_{i=1}^{c}\sum_{j=1}^{n}(C_{FCM}(u_{ij})^{m_1} + C_{PCM}(t_{ij})^{\theta_2})\|x_j - v_i\|_A^2 + \sum_{i=1}^{c}\eta_i\sum_{j=1}^{n}(t_{ij}^{\theta_2}\ln(t_{ij}^{\theta_2}) - t_{ij}^{\theta_2}) \\ J_{m_2,\theta_1}(U,T,V) = \sum_{i=1}^{c}\sum_{j=1}^{n}(C_{FCM}(u_{ij})^{m_2} + C_{PCM}(t_{ij})^{\theta_1})\|x_j - v_i\|_A^2 + \sum_{i=1}^{c}\eta_i\sum_{j=1}^{n}(t_{ij}^{\theta_1}\ln(t_{ij}^{\theta_1}) - t_{ij}^{\theta_1}) \\ J_{m_2,\theta_2}(U,T,V) = \sum_{i=1}^{c}\sum_{j=1}^{n}(C_{FCM}(u_{ij})^{m_2} + C_{PCM}(t_{ij})^{\theta_2})\|x_j - v_i\|_A^2 + \sum_{i=1}^{c}\eta_i\sum_{j=1}^{n}(t_{ij}^{\theta_2}\ln(t_{ij}^{\theta_2}) - t_{ij}^{\theta_2}) \end{cases} \quad (10)$$

Similar to EPFCM, by differentiating these equations and setting them to zero, we can obtain the updating formulas. The primary lower and upper fuzzy memberships $[\underline{u}_{ij}, \overline{u}_{ij}]$ are determined as follows:

$$\overline{u}_{ij} = max\left[\left(\sum_{t=1}^{c}\left(\frac{D_{ij}}{D_{tj}}\right)^{\frac{2}{m_1-1}}\right)^{-1}, \left(\sum_{t=1}^{c}\left(\frac{D_{ij}}{D_{tj}}\right)^{\frac{2}{m_2-1}}\right)^{-1}\right] \quad (11)$$

$$\underline{u}_{ij} = min\left[\left(\sum_{t=1}^{c}\left(\frac{D_{ij}}{D_{tj}}\right)^{\frac{2}{m_1-1}}\right)^{-1}, \left(\sum_{t=1}^{c}\left(\frac{D_{ij}}{D_{tj}}\right)^{\frac{2}{m_2-1}}\right)^{-1}\right] \quad (12)$$

Also, the primary lower and upper possibilistic typicality $[\underline{t}_{ij}, \overline{t}_{ij}]$ are as follows:

$$\overline{t}_{ij} = max\left[\left(exp\left(-W\left(\frac{\theta_1^2(\theta_1^2-1)C_{PCM}D_{ij}^2}{\eta_i}\right)\right)\right)^{\frac{\theta_1}{\theta_1^2-1}}, \left(exp\left(-W\left(\frac{\theta_2^2(\theta_2^2-1)C_{PCM}D_{ij}^2}{\eta_i}\right)\right)\right)^{\frac{\theta_2}{\theta_2^2-1}}\right], (13)$$

$$\underline{t}_{ij} = min\left[\left(exp\left(-W\left(\frac{\theta_1^2(\theta_1^2-1)C_{PCM}D_{ij}^2}{\eta_i}\right)\right)\right)^{\frac{\theta_1}{\theta_1^2-1}}, \left(exp\left(-W\left(\frac{\theta_2^2(\theta_2^2-1)C_{PCM}D_{ij}^2}{\eta_i}\right)\right)\right)^{\frac{\theta_2}{\theta_2^2-1}}\right]. (14)$$

To decrease the free parameters of our model and also to put the final uncertainty into the range of 0 to 1, the following constraints are added to the problem:

$$0 \leq C_{FCM} \leq 1, \quad 0 \leq C_{PCM} \leq 1, \quad C_{FCM} + C_{PCM} = 1.$$

Now, we have two upper bounds and two lower bounds in the problem, but we need one upper bound and one lower bound to calculate the centroids of the clusters straightforwardly. For this purpose, the following equations are presented for calculating the final upper and lower uncertainty representations:

$$\begin{cases} \omega_{ij}^1 = C_{FCM}.\overline{u}_{ij} + C_{PCM}.\overline{t}_{ij} \\ \omega_{ij}^2 = C_{FCM}.\overline{u}_{ij} + C_{PCM}.\underline{t}_{ij} \\ \omega_{ij}^3 = C_{FCM}.\underline{u}_{ij} + C_{PCM}.\overline{t}_{ij} \\ \omega_{ij}^4 = C_{FCM}.\underline{u}_{ij} + C_{PCM}.\underline{t}_{ij} \end{cases},$$

$$\begin{cases} \overline{\omega}_{ij} = max\,(\omega_{ij}^1, \omega_{ij}^2, \omega_{ij}^3, \omega_{ij}^4) \\ \underline{\omega}_{ij} = min\,(\omega_{ij}^1, \omega_{ij}^2, \omega_{ij}^3, \omega_{ij}^4) \end{cases}. \quad (15)$$

The centroids of clusters in the IT2EPFCM are calculated using the fuzzy and possibilistic memberships. Also, these memberships are defined based on IT2 FSs. Consequently, each centroid is represented by an interval between $v^L$ and $v^R$. To obtain the centroid of the proposed model, type reduction and defuzzification methods are required. Type reduction refers to the process of mapping a T2 FSs to T1 FSs (Wu & Mendel, 2009).

Among the various choices of type reduction methods, the Enhanced Karnik-Mendel (EKM) algorithm is used to estimate the cluster centers without using all of the embedded fuzzy sets. Details on this method can be found in (Wu & Mendel, 2009). After estimating the maximum and minimum values of cluster centers, the estimated interval type-1 fuzzy set fuzzy set for the cluster centers can be calculated as:

$$\tilde{v}_i = 1.0/[v_i^L, v_i^R].$$

Finally, a defuzzification method is used to estimate crisp centers. This can be computed as follows:

$$v_i = \frac{v_i^L + v_i^R}{2}.$$

Since the pattern set is extended to an interval type-2 fuzzy set, we should estimate the cluster centers using the hard partitioning of the fuzzy memberships. Moreover, the type of interval type-2 fuzzy set should be reduced before hard-partitioning. To achieve this aim, the left and right memberships ($\omega_{ij}^L, \omega_{ij}^R$) for all patterns are calculated by considering the left and right centroids ($v_i^L, v_i^R$). Then type-reduction is performed using the following equation:

$$\omega_{ij} = \frac{\omega_{ij}^L + \omega_{ij}^R}{2}, \quad i = 1, \ldots, c\,; j = 1, \ldots, n.$$

Eventually, hard partitioning can be obtained using the following rule:

**For** $i \neq k, i = 1, \ldots, c$ and $k = 1, \ldots, c$  **If**  $\omega_{ij} > \omega_{kj}$  **Then** assign $x_j$ to cluster $i$.

Based on the above discussions, the main steps of IT2EPFCM clustering method can be summarized as follows:

---

**Algorithm 2:** IT2 EPFCM clustering method

---

**Initialization:**

Fix $c, K, m_1, m_2, \theta_1, \theta_2, \epsilon, C_{FCM}, C_{PCM}$ and $I_{Max}$

Set iteration counter $t = 1$;

Initialize $\overline{\mathbf{U}}, \underline{\mathbf{U}}, \overline{\mathbf{T}}, \underline{\mathbf{T}}, \mathbf{\eta}, \overline{\mathbf{\omega}}$, and $\underline{\mathbf{\omega}}$ randomly.

**While** ($\left\|\overline{\mathbf{\omega}}^{(t)} - \overline{\mathbf{\omega}}^{(t-1)}\right\| > \epsilon$  or  $\left\|\underline{\mathbf{\omega}}^{(t)} - \underline{\mathbf{\omega}}^{(t-1)}\right\| > \epsilon$  or  $t < I_{Max}$)

   Update $\overline{\mathbf{U}}$ and $\underline{\mathbf{U}}$ using (11) and (12), respectively.

   Update $\overline{\mathbf{T}}$ and $\underline{\mathbf{T}}$ using (13) and (14), respectively.

   Update $\overline{\mathbf{\omega}}$ and $\underline{\mathbf{\omega}}$ using (15).

   $t = t + 1$

**End While**

**Return** cluster centers and membership values after performing type reduction and defuzzification

---

A remarkable feature of the proposed method is that by using the different values for its parameters ($\theta_1, \theta_2, m_1, m_2$, $C_{FCM}$ and $C_{PCM}$) it can behave like the conventional FCM, PCM, IFCM, IPCM, and PFCM methods; as a result, the proposed method can be used as a general approach to fuzzy clustering. Furthermore, in the Zexuan, et al. model, the coincident-clusters problem still exists, but it has been corrected our algorithms. However, since the main goal of this study is to enhance PFCM; from now on, we just focus on EPFCM and IT2EPFCM.

## 5. Experimental results

In this section, several examples are presented to demonstrate the performance of the proposed algorithms. We test and compare the results of the proposed methods with those obtained by FCM, PCM, PFCM, IT2FCM, IT2PCM, and IT2PFCM. The results of these experiments are evaluated using three performance criteria, namely Reconstruction Error (RE), Rand Index (RI)(Rand, 1971), and three fuzzy cluster validity indices.

In possibilistic and fuzzy clustering analysis, every point has a degree of belonging to the centroid of clusters. Furthermore, each point can be regenerated using the center of clusters and the degree of belonging to the clusters. Consequently, the reconstruction error for a data point like $x_i$ is computed as the distance between $x_i$ and its regenerated version $x_i'$.

Rand Index (RI) is a measure of the similarity between two clustering results (e.g., predicted labels $\hat{y}$ and true labels $y$). It estimates the likelihood of an element being correctly classified. RI index can be computed as follows:

$$RI = \frac{\alpha + \beta}{N(N-1)/2}$$

where $\alpha$ denotes the number of times a pair of elements belongs to a same cluster across $\hat{y}$ and $y$; $\beta$ is the number of times a pair of elements are in different clusters across $\hat{y}$ and $y$; and $N$ is the number of elements. This index ranges from 0 to 1, with a higher value representing a better clustering result.

Furthermore, cluster validity indices (CVI) will be used for evaluating the fitness of partitions produced by clustering algorithms. Once the partitions are obtained by a clustering method, the validity function can help us to validate whether it accurately presents the data structure or not. For that purpose, three well-known validity indices including Fukuyama and Sugeno (FS), Xie and Beni (XB), and Kowan are used to compare the performance of the proposed algorithms to other baseline methods. For these CVIs, the lower index value for an algorithm indicates the better performance of that algorithm. For more detail on the concept of fuzzy validity indices and the techniques directly relevant to our work, please refer to (Fazel Zarandi et al., 2021; Wang & Zhang, 2007).

In all experiments, the process of clustering and calculating the performance metrics has been repeated 5 times to verify the precision of our final result. Average of the obtained performance metrics are reported in the tables.

For all experiments we use the following computational protocols: $\varepsilon = 0.0001$ and the maximum number of iterations is 100, $C_{FCM} = 0.8$, and $C_{PCM} = 0.2$. Besides, for type-1 fuzzy and possibilistic clustering algorithms, we set $m = 1.5$, and $\theta = 6$. For type-2 fuzzy and possibilistic clustering algorithms, we set $m_1 = 1.1$, $m_2 = 1.5$, $\theta_1 = 1.5$, and $\theta_2 = 3$. Moreover, all the experiments are performed on a computer with Intel Core i7 CPU 3.1 GHz processor, 10 GB of RAM, using Microsoft Windows 10 (64 bit) OS.

**Example 1.** In the first experiment, EPFCM is compared to other T1 fuzzy clustering methods. To that end, $D_2$ consists of 520 data points and seven clusters is used. This data set is demonstrated in Figure 2.

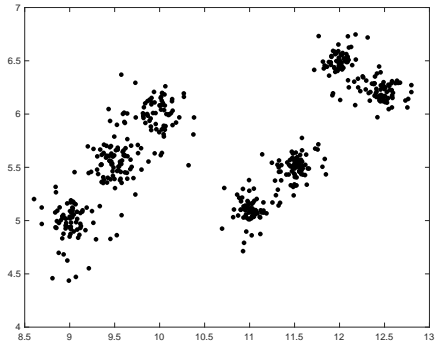

**Figure 2.** Scatter plot of $D_2$ data set containing seven clusters.

RE, Kowan, and FS are computed for this dataset using FCM, PCM, PFCM and EPFCM algorithms and are presented in Table 1. As can be seen, the proposed EPFCM clustering outperforms other T1 fuzzy clustering algorithms. Moreover, PFCM works well and is the second-best method.

**Table 1.** REs and CVIs of clustering results obtained by performing the T1 fuzzy clustering algorithms.

| Performance metric | FCM | PCM | PFCM | EPFCM |
|---|---|---|---|---|
| RE | 0.0092 | 0.0095 | 0.0088 | **0.0080** |
| FS | -838.73 | -588.39 | -828.59 | **-845.63** |
| Kowan | 805.89 | 980.66 | 480.17 | **299.61** |

**Example 2.** In the second experiment, the proposed IT2EPFCM is compared to IT2FCM, IT2PCM, and IT2PFCM. This example involves four generated corner-shape clusters ($D_3$ data set) consists of 1600 data points as shown in Figure 3 (a). In this experiment, noisy points are added to this data set to better evaluate the robustness of our methods in the presence of noise. For this purpose, 50, 100, and 200 normally distributed noisy points were added to this dataset (Figure 3(b-d)). The obtained REs and XBs are reported in Table 2. According to our computations, the proposed algorithm is less sensitive to noise and the problems with a high level of uncertainty are better handled by the proposed interval type-2 clustering method. Additionally, IT2PFCM shows a reasonably good overall performance and is the second-best method. However, the differences between IT2EPFCM's performance and IT2PFCM is significantly higher for $D_3$ with noisy points. This fact demonstrates the robustness of our method in the presence of noise.

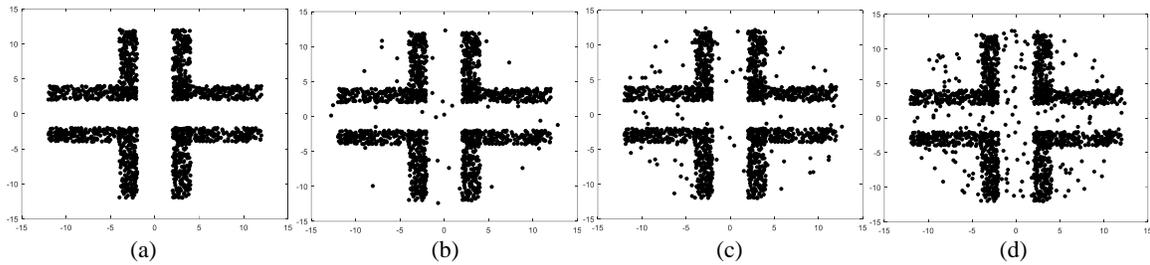

**Figure 3.** Scatter plot of $D_3$ data set containing four corner-shape clusters (a) without noisy points (b) $D_3$+50 noisy points (c) $D_3$+100 noisy points (d) $D_3$+200 noisy points.

**Table 2.** REs and XBs of clustering results obtained by performing the interval type 2 clustering algorithms.

| Data set | | IT2FCM | IT2PCM | IT2PFCM | IT2EPFCM |
|---|---|---|---|---|---|
| $D_3$ | RE | 0.1315 | 0.1572 | 0.1155 | **0.1138** |
| | XB | 87.45 | 111.95 | 106.52 | **78.69** |
| $D_3$+50 noise points | RE | 0.1235 | 0.1526 | 0.1198 | **0.1157** |
| | XB | 235.01 | 255.75 | 271.11 | **197.91** |
| $D_3$+100 noise points | RE | 0.1194 | 0.1541 | 0.1184 | **0.1059** |
| | XB | 106.23 | 135.81 | 99.54 | **90.45** |
| $D_3$+200 noise points | RE | 0.1220 | 0.1486 | 0.1107 | **0.1087** |
| | XB | 145.32 | 198.34 | 86.12 | **66.68** |

**Example 3.** For our next experiment, three high dimensional data, namely iris plants database (Iris), Breast Cancer Coimbra, and Yeast with a known number of clusters are used to evaluate the performance of the proposed algorithms. To that end, RI is used to compare the proposed algorithm to their T1 and T2 counterparts. All the datasets are adopted from UCI Machine Learning Repository (Dua & Graff, 2019; Patrício et al., 2018), and the details of them are given in Table 3 where r , N , and c indicate the number of features, number of data vectors, and number of clusters, respectively.

Table 3. The details of the high dimensional data sets.

| Data set | r | N | c |
|---|---|---|---|
| Iris | 4 | 150 | 3 |
| Breast Cancer Coimbra | 9 | 116 | 2 |
| Yeast | 8 | 1484 | 10 |

The values of the Rand index for all algorithms are reported in Figure 4. As can be seen, the RI values for the EPFCM algorithm are higher than PCM, FCM, PFCM, and even IT2PCM algorithms. Moreover, the proposed IT2EPFCM algorithm performs steadily better than IT2PFCM clustering method. We can clearly see that the proposed IT2EPFCM performed the best among all methods.

It worth mentioning that the only goal of this example is to compare the T1 and T2 fuzzy clustering algorithms under the same experimental settings. It is definitely possible to achieve higher RI using these clustering algorithms, but since we want to have a fair comparison, we did not tune the parameters of these algorithms.

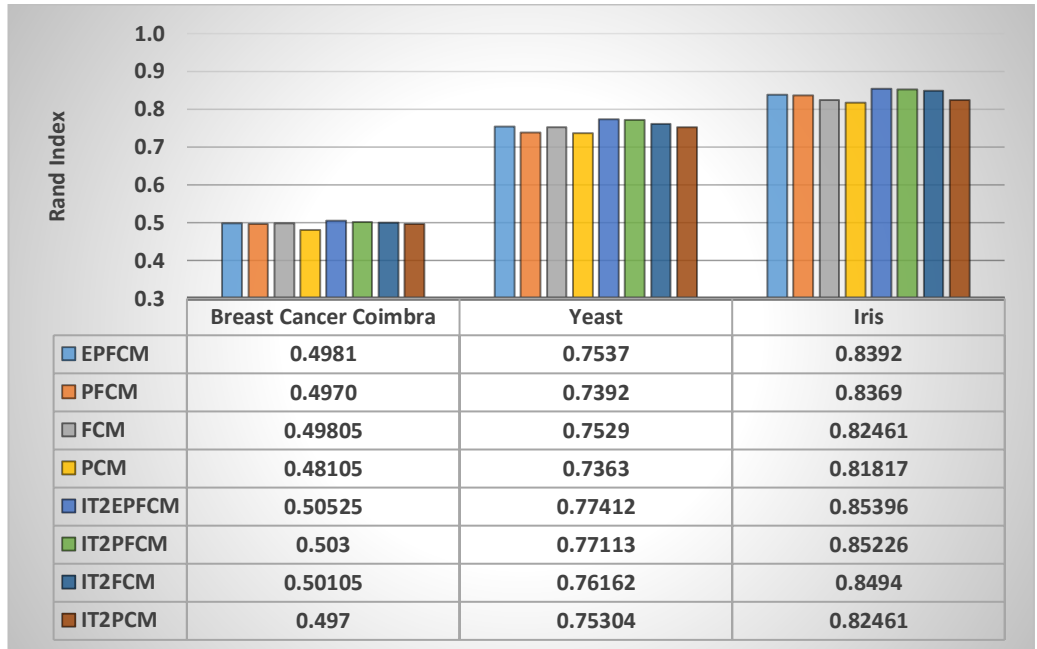

| | Breast Cancer Coimbra | Yeast | Iris |
|---|---|---|---|
| EPFCM | 0.4981 | 0.7537 | 0.8392 |
| PFCM | 0.4970 | 0.7392 | 0.8369 |
| FCM | 0.49805 | 0.7529 | 0.82461 |
| PCM | 0.48105 | 0.7363 | 0.81817 |
| IT2EPFCM | 0.50525 | 0.77412 | 0.85396 |
| IT2PFCM | 0.503 | 0.77113 | 0.85226 |
| IT2FCM | 0.50105 | 0.76162 | 0.8494 |
| IT2PCM | 0.497 | 0.75304 | 0.82461 |

Figure 4. The values of the Rand Index for three high dimensional data sets.

**Example 4.** To evaluate the computational cost of proposed algorithms, we compare their computational time costs to baseline methods. For that purpose, we use 3 datasets, namely $D_4$, Iris, and Wine data sets. $D_4$ is consists of 140 data points and 2 clusters and is demonstrated in Figure 5. Moreover, Wine dataset is also adopted from UCI Machine

Learning Repository (Dua & Graff, 2019). We perform each experiment 10 times with randomly initialized cluster centers. The average execution time and the number of iterations (NI) are presented in Table 4. Both the number of iterations and the runtime of EPFCM are slightly higher, but still comparable with those in the other algorithms (PCM and PFCM). Moreover, the runtime of IT2EPFCM is comparable with that of other IT2 algorithms. Also, it can generally reach convergence with fewer iterations.

In general, the proposed algorithms have slightly higher running time compared with all other algorithms. However, as we saw in the previous experiments, they outperform their counterparts. Thus, slightly higher runtime is a small price to pay for obtaining satisfactory results and higher accuracy. Furthermore, comparing to other algorithms, they can reach convergence with fewer iterations.

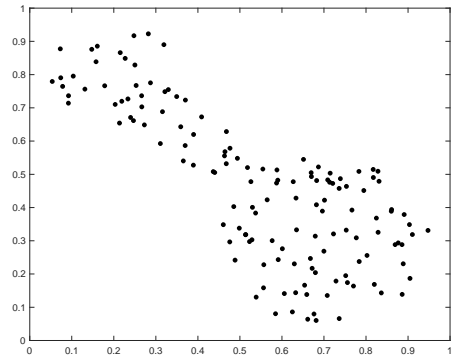

**Figure 5.** Scatter plot of $D_4$ data set containing two clusters.

**Table 4.** Average of Runtime (seconds) and the number of iterations (mean±standard deviation) for various clustering algorithms.

| Method | $D_4$ | | Iris | | Wine | |
|---|---|---|---|---|---|---|
| | **Runtime** | **NI** | **Runtime** | **NI** | **Runtime** | **NI** |
| FCM | 0.0021 | 13.4 ± 1.2 | 0.0032 | 24.5 ± 4.3 | 0.0019 | 11.4 ± 0.3 |
| PCM | 0.0051 | 43.8 ± 3.7 | 0.0037 | 13.2 ± 0.5 | 0.0028 | 10.5 ± 1.2 |
| PFCM | 0.0063 | 12.2 ± 3.1 | 0.0077 | 42.1 ± 4.4 | 0.0044 | 11.1 ± 0.7 |
| EPFCM | 0.0071 | 16.3±2.7 | 0.0062 | 18.2±2.4 | 0.0051 | 10.2±2.1 |
| IT2FCM | 0.0163 | 8.3 ± 0.9 | 0.0099 | 8.9±1.6 | 0.0113 | 9.6±1.6 |
| IT2PCM | 0.165 | 8.9 ± 1.5 | 0.0096 | 8.7±1.9 | 0.0132 | 9.3 ± 1 |
| IT2PFCM | 0.0178 | 7.9 ± 1.4 | 0.0119 | 9.1±1.4 | 0.0137 | 9.4 ± 0.8 |
| IT2EPFCM | 0.0185 | 7.2±1.9 | 0.0133 | 8.3±0.7 | 0.0135 | 9.7±0.5 |

## 6. Application in gene expression data analysis

Clustering algorithms have demonstrated their potential to find the underlying patterns in microarray gene expression profiles. In this section, two microarray gene expression datasets, namely Rat CNS and Arabidopsis thaliana are used and the capability of the proposed algorithms will be analyzed from various perspectives. These datasets are adopted from (Maulik et al., 2009). Arabidopsis Thaliana dataset contains expression levels of 138 genes of Arabidopsis Thaliana over 8 time points. The Rat CNS dataset is comprised of the expression levels of a set of 112 genes during rat central nervous system development. It has 9 dimensions each of them representing 9 time points. The number of clusters for these datasets is computed using FP validity index (Fazel Zarandi et al., 2021). Based on our computations, the near-optimal number of clusters for Arabidopsis dataset is 4 clusters and for Rat CNS dataset is 3 clusters.

In the first step, to visually observe the clustering results, we apply EPFCM algorithm and Eisen plots of the clustered datasets are presented. We also generated a random sequence of genes for the simpler distinction between data before and after clustering. These plots are demonstrated in Figures 6 and 7. In these figures, bright colors denote higher expression levels while dark colors denote lower expression values. Furthermore, the red lines in these plots are the boundaries of clusters.

Needless to say, the expression profiles of the genes in each cluster are similar to each other and they have similar color patterns. In other words, EPFCM can put similar genes beside each other. Also, the genes in different clusters are properly separated. In the second step, we use Silhouette index (Rousseeuw, 1987) to compare the performance of the proposed algorithms to their counterparts. This index evaluates how well an observation is clustered, and it can be defined as follows:

$$S = \frac{b - a}{\max(b, a)},$$

where $a$ is the average distance of a point from all other points in the same cluster, and $b$ is the average distances of the point from all other points in the closest cluster. This index ranges from -1 to 1, with a higher value representing a better clustering result. We compare the methods proposed here to eleven state-of-the-art approaches for gene expression clustering including centroid linkage (UPGMC), weighted average linkage (WPGMA), unweighted average linkage (UPGMA), gaussian mixture model (GMM), partitioning around medoids (PAM), self-organizing map (SOM), k-means, general type-2 possibilistic fuzzy c-means (GT2PFCM), IT2PFCM, PFCM, and FCM. We chose PFCM and IT2PFCM because these clustering methods demonstrated a really good performance based on previous experiments. For more information about these algorithms, please refer to (Ji et al., 2014; Kerr et al., 2008; Oyelade et al., 2016).

The process of clustering and calculating Silhouette index has been repeated 25 times to verify the precision of our final results. The averages of the obtained results are reported in Figure 8. As can be seen, IT2EPFCM outperforms its counterparts in both gene expression datasets. Interestingly, it is even better than GT2PFCM that is the result of using EPFCM instead of PFCM in its objective function. Moreover, if we ignore T2 fuzzy algorithms, EPFCM achieves a better or comparable performance compared with other algorithms. Generally, the fuzzy algorithms demonstrate superior performance over the hard-clustering algorithms.

It should be noted that since we want to keep the main models as simple as possible, we used simple methods for type reduction and defuzzification steps. These steps can have a huge impact on the clustering results. Thus, using more sophisticated methods, we may be able to further improve our results. For more information about these methods, please refer to (Torshizi et al., 2015).

Eventually, to gain a better insight into IT2EPFCM's clustering results, the cluster profile plots of these datasets are presented in Figures 9 and 10. The x-axis denotes the 8 or 9 time points that data were collected, and y-axis denotes

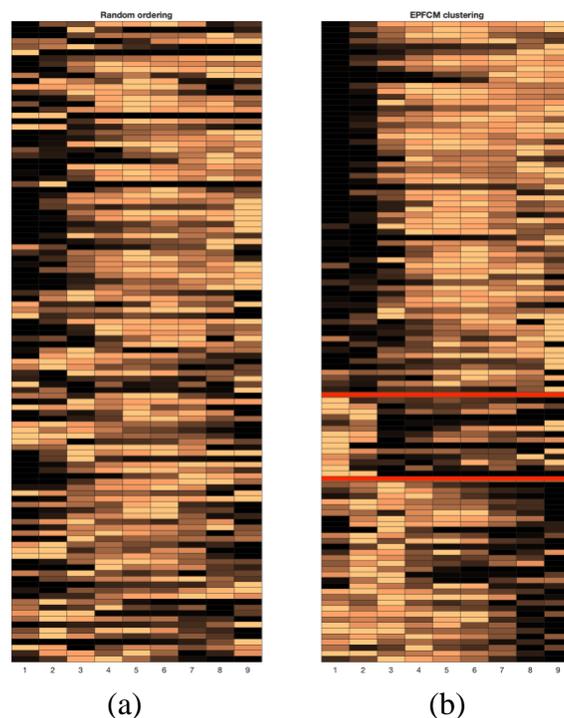

(a) (b)
**Figures 6**. Eisen plot for Rat CNS (a) random sequence of genes (b) clustered data.

the expression values of genes in each cluster. Obviously, IT2EPFCM is able to generate compact well-separated clusters. In other words, the general form of profile plots within each cluster are similar to each other while the profile plots from different clusters are dissimilar.

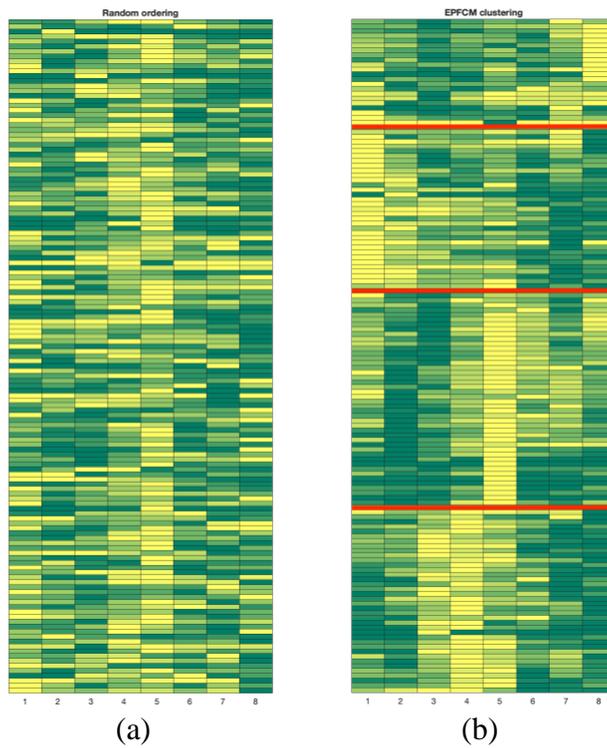

(a)          (b)

**Figures 7**. Eisen plot for Arabidopsis thaliana (a) random sequence of genes (b) clustered data.

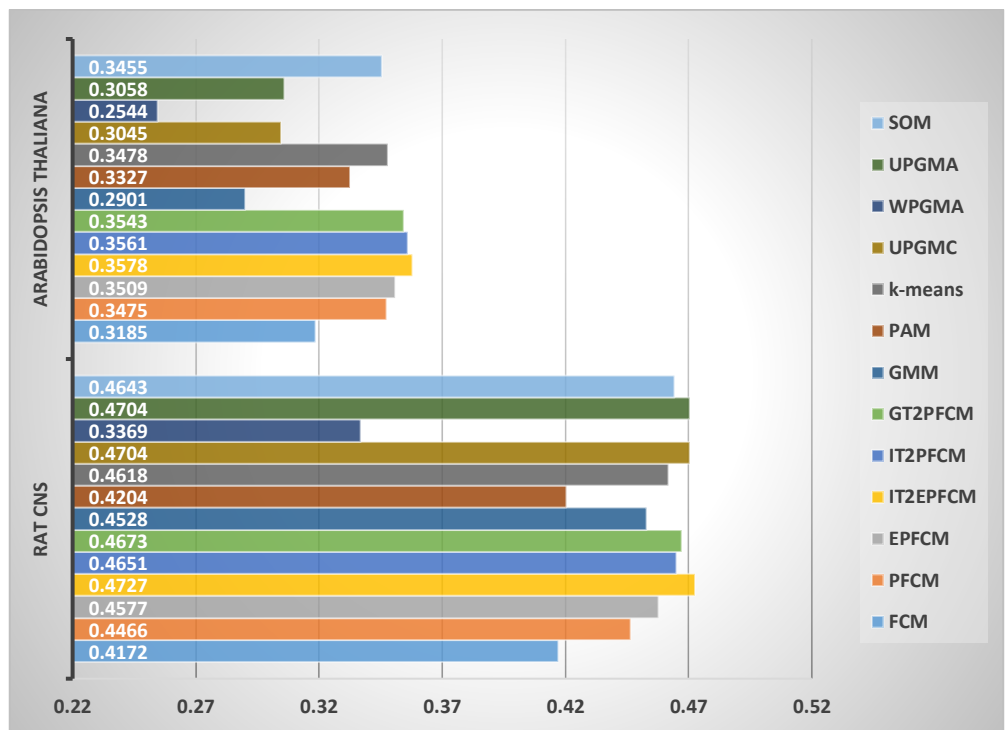

**Figures 8**. The average of Silhouette index values for various clustering algorithms.

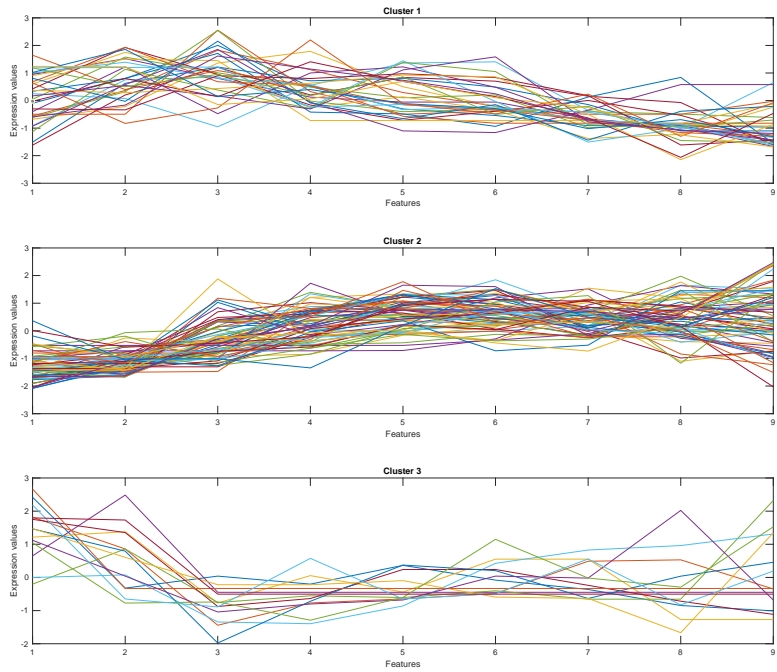

**Figures 9**. Cluster profile plots for Rat CNS.

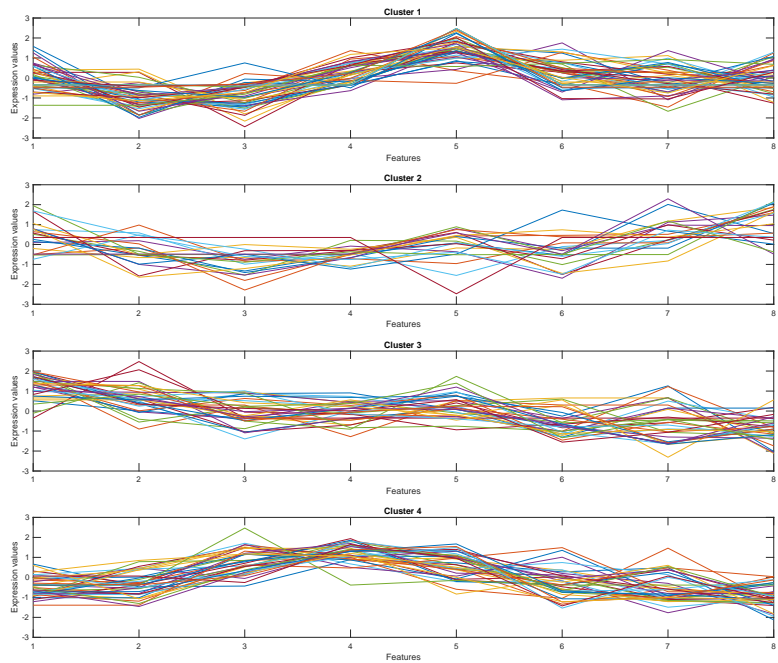

**Figures 10**. Cluster profile plots for Arabidopsis thaliana.

## 7. Conclusions and future work

Even though PFCM clustering performs better in the presence of noise over conventional FCM, cluster coincidence still occurs in PFCM algorithm. We analyzed the problem of cluster coincidence in the PCM-based algorithms (i.e. conventional PFCM and IT2PFCM) and solved this issue by presenting EPFCM method. Then, interval type-2 enhanced possibilistic fuzzy c-means clustering models, IT2EPFCM, introduced based on parameter uncertainties. In IT2EPFCM algorithm, we consider two fuzzifiers $(m_1, m_2)$ for fuzzy memberships and two fuzzifiers $(\theta_1, \theta_2)$ for possibilistic typicalities to express the uncertainty associated with the fuzzifier parameters of the possibilistic fuzzy clustering.

To provide a comprehensive performance analysis, the proposed approaches evaluated using different artificial and real-world data sets. The experiments show that our methods can solve the problems of FCM, PCM, and PFCM algorithms. Furthermore, our results demonstrated the superiority and flexibility of the proposed algorithms under highly uncertain environments compared to several existing clustering algorithms in the literature.

Eventually, we applied the proposed algorithms to gene expression data clustering. The comparative analysis illustrated that IT2EPFCM model outperforms the best of its counterparts and yields higher performance.

Even though our model is quite promising, it also suffers from some limitations that can be addressed in future works. As discussed earlier, one of the outstanding features of IT2 FSs is their resilience to noise compared to T1 FSs. This resilience can be enhanced by introducing GT2 FSs. Second, the computational complexity associated with the T2 clustering algorithms is an obstacle for some applications. As future work, we will develop more efficient algorithms using bio-inspired optimization techniques.

Furthermore, an interesting future direction is to extend the proposed framework to a semi-supervised scenario. Finally, we can extend the proposed models to deal with more complex data, e.g., data with missing entries.